\documentclass[12pt]{article}
\usepackage{amsmath}
\usepackage{amssymb}
\usepackage{boxedminipage}
\parskip=\medskipamount
\usepackage{boxedminipage}

 1   
\font\ddpp=msbm10  scaled \magstep 1  

\def\QED{\hskip0.1em\hfill\null\ \null\nobreak\hfill
\kern3pt\lower1.8pt\vbox{\hrule\hbox
{\vrule\kern1pt\vbox{\kern1.7pt \hbox{$\scriptstyle
QED$}\kern0.2pt}\kern1pt\vrule}\hrule}}

\def\R{\hbox{\ddpp R}}               

\numberwithin{equation}{section}

\def\beqn{\begin{equation}}
\def\eeqn{\end{equation}}

\begin{document}

\title{{\Large \textbf{A unified geometric treatment of material defects}}}

\author{{\ \textsc{Marcelo Epstein}} \\
{\small University of Calgary, Canada }\\
{\small mepstein@ucalgary.ca }\\ \\
{\ \textsc{Reuven Segev}} \\
{\small Ben-Gurion University of the Negev, Israel }\\
{\small rsegev@bgu.ac.il}\\}
\date{} \maketitle

\begin{abstract}
A unified theory of material defects, incorporating both the smooth and the singular descriptions, is presented based upon the theory of currents of Georges de Rham. The fundamental geometric entity of discourse is assumed to be represented by a single differential form or current, whose boundary is identified with the defect itself. The possibility of defining a less restrictive dislocation structure is explored in terms of a plausible weak formulation of the theorem of Frobenius. Several examples are presented and discussed.
\end{abstract}

\bigskip
{\bf Keywords}: Dislocations --- de Rham currents --- differential geometry --- Frobenius theorem --- Frank's rules --- integrability --- distributions
\bigskip

\section*{Introduction}\label{sec:introduction}
While the precise definition of the concept of material defect must be left to every particular context, a common feature of all theories dealing with defects (dislocations, inhomogeneity, and so on) appears to be that the presence of defects translates itself mathematically into the lack of integrability of some geometric entity. In a general differential geometric framework, questions of integrability pertain to differential forms and their exactness or lack thereof. It seems appropriate, therefore, to undertake a unified treatment of defects by associating to any possible structure under consideration one or more differential forms. On the other hand, since differential forms are, by definition, smooth entities, it would appear that the rich variety of isolated defects, whose practical and historical importance cannot be denied, might be left out and that a unified treatment encompassing both the continuous and discrete cases would remain out of reach of a single formal apparatus. The situation is similar in many other engineering applications, where concentrated entities (forces, masses, charges) can be seen as limiting cases of their smooth counterparts. The unified mathematical treatment of these cases was historically achieved by the theory of distributions, where the singular entities are represented not by functions but rather by linear functionals on a suitable space of test functions. The most common example is provided by the Dirac delta which assigns to each compactly supported function in $\R$ its value at the origin. Since a scalar field can be considered as a particular case of a differential form, it is not surprising that L. Schwartz's theory of distributions can be extended to forms of all orders. This extension, achieved by G. de Rham \cite{derham}, is completely general and independent of any metric considerations, a feature that should be considered essential in a truly general geometric setting. De Rham introduced the terminology of {\it currents} to designate his generalized differential forms. It is this tool that will serve our purposes in the present formulation of the unified theory of defects.\footnote{This work extends our previous article \cite{epseg}.}

\section{Currents} \label{sec:currents}
\subsection{Definition} \label{sec:definition}
A $p$-current on an $n$-dimensional manifold $\mathcal M$ is a continuous\footnote{By continuity we mean that the sequence of evaluations $T[\phi_i]$ on a sequence of $C^\infty$ $p$-forms supported within a common compact subset of $\mathcal M$ tends to zero whenever the coefficients and all their derivatives of a coordinate representation of the forms $\phi_i$ tend to zero uniformly as $i \to \infty$.} linear functional $T[\phi]$ on the vector space of all $C^\infty$ $p$-forms $\phi$ with compact support in $\mathcal M$. To understand in what sense this definition is consistent with that of smooth forms, it suffices to exhibit the latter as a particular case of the former. Let, therefore, $\omega$ represent a smooth $p$-form on $\mathcal M$. We can uniquely associate to it the ($n-p$)-current $T_\omega$ defined as the linear operator
\begin{equation}\label{eq1}
T_\omega[\phi]=\int\limits_{\mathcal M} \omega \wedge \phi,
\end{equation}
for all ($n-p$)-forms $\phi$ with compact support in $\mathcal M$. Strictly speaking, the ($n-p$)-current $T_\omega$ cannot be ``equal'' to the $p$-form $\omega$, but they are indistinguishable from each other in terms of their integral effect on all ``test forms'' $\phi$. Thus, a form bears to its associated current the same relation that a function bears to its associated distribution. An important non-trivial example of a current that is not associated to any differential form is the following. Let $s$ be a $p$-simplex in $\mathcal M$. We associate to it the $p$-current defined by:
\begin{equation} \label{eq2}
T_s[\phi]=\int\limits_s \phi,
\end{equation}
for all compactly supported $p$-forms $\phi$. The definition above can be extended by linearity to arbitrary chains. These examples show how an integrand and a domain of integration are unified under the single formal umbrella of currents.
\subsection{Operations} \label{sec:operations}
\begin{enumerate}
\item Currents of the same dimension can be added together and multiplied by real numbers in an obvious way.
\item The product of a $p$-current $T$ with a $q$-form $\alpha$ is the ($p-q$)-current $T\llcorner\alpha$ defined as:
\begin{equation} \label{eq3}
(T \llcorner \alpha)[\phi]=T[\alpha\wedge\phi].
\end{equation}
Similarly,
\begin{equation} \label{eq4}
\alpha \lrcorner T= (-1)^{(n-p)q}\; T\llcorner\alpha.
\end{equation}
\item The product of a $p$-current $T$ with a vector field $\bf X$ is the ($p+1$)-current:
\begin{equation} \label{eq5}
(T\wedge {\bf X})[\phi] = T({\bf X}\llcorner \phi].
\end{equation}
\item The boundary of a $p$-current $T$ is the ($p-1$)-current
\begin{equation} \label{eq6}
\partial T [\phi] = T [d \phi].
\end{equation}
\end{enumerate}
Using Stokes' theorem for chains, it is easy to show that, for any chain $c$,
\begin{equation} \label{eq7}
\partial T_c = T_{\partial c}.
\end{equation}

\subsection{Possibilities} \label{sec:possibilities}
The notion of currents opens the doors for generalizing classically smooth differential geometric objects, such as connection, torsion and curvature. While this idea is beyond the scope of this paper, it is not difficult to intuit the possibilities. Consider, for example, a (non-zero) decomposable differential $p$-form $\omega$ on an $n$ dimensional differentiable manifold $\mathcal M$. Thus, there exist $p$ linearly independent 1-forms $\omega_i\;(i=1,...,p)$ such that:
\begin{equation} \label{eq8}
\omega=\omega_1 \wedge ...\wedge \omega_p.
\end{equation}
Such a form uniquely determines at each point of $x \in \mathcal M$ an ($n-p$)-dimensional subspace $H_x$ of the tangent space $T_x{\mathcal M}$. A vector ${\bf v} \in T_x{\mathcal M}$ belongs to $H_x$ if $\omega_i({\bf v}) = 0$ for each $i=1,...,p$. The collection $\mathcal H$ of all the subspaces $H_x$ is called a (geometric) ($n-p$)-dimensional {\it distribution} on $\mathcal M$. Conversely, given a distribution, the corresponding decomposable form $\omega$ is determined up to multiplication by a scalar field $\alpha: {\mathcal M} \to {\R}$. A submanifold $\mathcal S$ of $\mathcal M$ is called an {\it integral manifold} of the distribution $\mathcal H$ if for every $s \in {\mathcal S}$ we have $T_s {\mathcal S} = H_s$. A distribution is {\it completely integrable} if at every point $x$ it admits an integral manifold of maximal dimension (i.e., $n-p$). According to (one of the versions of) the theorem of Frobenius, a distribution $\mathcal H$ defined by a decomposable form $\omega$ is completely integrable if, and only if, there exists a 1-form $\beta$ such that:
\begin{equation} \label{eq9}
d \omega = \beta \wedge \omega.
\end{equation}
 This is tantamount to saying that, for some choice of the scalar degree of freedom $\alpha$, the form $\alpha \omega$ is closed, namely, there exists an integrating factor $\alpha$ such that $d(\alpha \omega) = 0$. So far, we have been dealing with the smooth case. Assume now that we have a means of characterizing the decomposability of a current (perhaps as the limit of a sequence of non-zero decomposable forms). We could now declare that a decomposable $p$-current $T$ determines a $p$-dimensional {\it singular geometric distribution} on $\mathcal M$ and define the complete integrability of the singular distribution by the condition:
 \begin{equation} \label{eq10}
 \partial T = \beta \lrcorner T,
 \end{equation}
 for some 1-form $\beta$.\footnote{As a curiosity, it is interesting to remark that Equation (\ref{eq10}) can be informally regarded as the eigenvalue problem of the boundary operator $\partial$. Its ``eigenvectors'' are the completely integrable currents.} A stronger condition would be to require that $\beta$ be closed. Notice that since connections in general can be regarded as (horizontal) distributions on fibre bundles, and since the curvature of a connection is related to its complete integrability, we can expect that singular connections can be introduced by means of decomposable currents and their non-vanishing curvature can be detected by the violation of a condition such as (\ref{eq10}). By this means, a situation is envisioned in which the standard curvature vanishes almost everywhere and is concentrated, as it were, at a single point. We remark that Equation (\ref{eq10}) is by no means the result of a theorem, but only a possible definition of complete integrability of a singular distribution. Clearly, the fact that $\beta$ is a smooth form may severely limit the singular distributions that can be considered completely integrable.

 \section{Bravais hyperplanes} \label{sec:bravais}
 \subsection{The smooth case} \label{sec:smooth}
 The traditional heuristic argument to introduce continuous distributions of dislocations in crystalline materials calls for the specification of a {\it frame field} (or {\it rep\`ere mobile}) in the body manifold, and the consequent distant parallelism.\footnote{This feature is present also, albeit with the degree of freedom afforded by material symmetries, in the constitutively based approach propounded by Kondo \cite{kondo} and Noll \cite{noll}, whereby points are compared, in a groupoid-like fashion, via material isomorphisms between their tangent spaces.} An alternative, dual, picture is obtained by means of a {\it co-frame field}, which can be regarded as an ${\R}^n$-valued 1-form on $\mathcal M$. This point of view suggests perhaps that $n$ linearly independent 1-forms might constitute a convenient point of departure for our desired generalization. Each of these 1-forms would represent a family of Bravais planes. It comes as a surprise, however, that defects are meaningful and detectable with just a single family of such planes or, more specifically and less surprisingly, that integrability conditions can be associated with a single 1-form on a manifold. Geometrically, a 1-form (always decomposable) induces an ($n-1$)-dimensional distribution, that is, a field of hyperplanes. It is physically important to point out that, relinquishing the multiplicative degree of freedom alluded to in the previous section, a 1-form also specifies a local {\it density} of these Bravais hyperplanes. Indeed, a covector acting on a vector space defines a family of parallel hyperplanes and the evaluation of the covector on a given vector can be pictorially regarded as the `number of hyperplanes' pierced by the vector. We have at our disposal, therefore, two somewhat different images induced by the specification of a 1-form $\omega$ on a manifold $\mathcal M$. The first one would look at the integrability of the form itself by demanding that $\omega$ be {\it closed}, namely:
 \begin{equation} \label{eq11}
 d \omega = 0.
 \end{equation}
 Physically, this condition means not only that the induced distribution is completely integrable, but also that the local hyperplane densities are mutually compatible. More to the point, we expect the Bravais hyperplanes to fit well with their neighbours not only as hyperplanes but also as stacks thereof at each point. The second image, on the other hand, would demand only the nice fit between the hyperplanes themselves. The corresponding (less demanding) integrability condition would take the form of a Frobenius condition. In either case we will refer to the form $\omega$ as a {\it layering form}.

 \subsection{Dislocations and currents}
 From the smooth case we learn two facts: (1) A basic entity to be analyzed for defectiveness is expressed in terms of a differential form $\omega$. In the case of a 1-form, the entity is a distribution of Bravais hyperplanes with their corresponding stacking densities. (2) The defectiveness of the structure is measured by the exterior derivative $\Omega=d\omega$ of the form representing the basic entity. It is natural, therefore, to identify $\Omega$ with the {\it dislocation} or, in more general terms, with the {\it imperfection}. Notice that different entities $\omega$ may have the same boundary (if they differ by a closed form), which means that the same dislocation structure may arise from different physical objects.

 The generalization of these notions to the non-smooth case is straightforward, provided one bears in mind de Rham's invention. Let a $p$-current $T$ represent some basic physical object. Then we call its boundary $D=\partial T$ the associated {\it dislocation current}. Notice that $D$ is a ($p-1$)-current. We say that the object represented by $T$ is defect-free if the integrability condition
 \begin{equation} \label{eq12}
 \partial T = 0,
 \end{equation}
 is satisfied. In other words, $T$ is defect-free if it has a vanishing boundary $D=0$. An important non-trivial example is provided by a $p$-simplex $s$ embedded in the manifold $\mathcal M$. As we have seen, we can associate to $s$ the current $T_s$ defined in Equation (\ref{eq2}). For the important particular case $p=n-1$, the current $T_s$ can be regarded as the specification of a Bravais structure concentrated on $s$ rather than distributed over the whole body. Physically, the simplex $s$ may be regarded as a cut inside the body where a putative layer of atoms has been inserted or removed. Let us calculate the corresponding dislocation current $D_s$. For every compactly supported ($p-1$)-form $\phi$, we have:
 \begin{equation} \label{eq13}
 D_s[\phi]=\partial T_s[\phi]=T_s[d\phi]=\int\limits_s d\phi=\int\limits_{\partial s} \phi = T_{\partial s} [\phi].
 \end{equation}
 Thus, we obtain the important result that the dislocation coincides with the current associated with the boundary of the embedded simplex.

 \subsection{An edge dislocation}
 Consider the open cube ${\mathcal M}=(-1,1)^3$ in ${\R}^3$ with coordinates $x,y,z$. Let $h$ denote its intersection with the (oriented) lower half-plane $x=0, z \le 0$. We associate with $h$ the current:
 \begin{equation} \label{eq13}
 T_h[\phi]=\int\limits_h \phi,
 \end{equation}
where $\phi$ denotes an arbitrary compactly supported 2-form in $\mathcal M$. Notice that the supports of these forms must be made entirely from (interior) points of the open cube. As a consequence of this observation, we obtain:
\begin{equation} \label{eq14}
D_h[\psi]=\partial T_h [\psi] = \int\limits_h d\psi = \int\limits_L \psi = T_L[\psi],
\end{equation}
where $L$ is the open interval $(-1,1)$ on the $y$-axis. Thus we recover the classical textbook description of an edge dislocation as the result of the removal of an atomic half plane.

\subsection{An open book} \label{sec:openbook}
An interesting example is provided by the 1-form $\phi=d\theta$ defined in $F={\R}^2\backslash\{0\}$, that is, the real plane devoid of the origin. We denote by $\rho, \theta$ the usual polar ``coordinates''. Clearly, these are not legitimate global coordinates for $F$ since ($\rho, \theta$) and ($\rho, \theta + 2\pi$) represent the same point. Nevertheless, the notation $d\theta$ is standard and reinforces the fact that the 1-form $\phi$ is closed (namely, $d\phi=0$) though not exact. If we should propose $\phi$ as a layering form on $F$, the corresponding distribution would look like the set of all rays emanating from (but not including) the origin $\{0\}$. The corresponding dislocation current is $D=d\phi=0$, which means that, as far as the set $F$ is concerned, the given layering is defect free. We are interested, however, in extending the form $\phi$ to include the missing origin. To this end, we define the following associated current on ${\R}^2$:
\begin{equation} \label{eq15}
T_\phi[\alpha]=\int\limits_F \phi \wedge \alpha.
\end{equation}
The subtle point in this definition is that the 1-forms $\alpha$ have compact support in ${\R}^2$ rather than in $F$. We are interested to obtain the corresponding dislocation current $D$, that is:
\begin{equation} \label{eq16}
D[\beta]=\partial T[\beta]=T[d\beta]=\int\limits_F \phi \wedge d\beta=\int\limits_F d(\beta \phi),
\end{equation}
where $\beta$ is any zero-form (function) with compact support in ${\R}^2$ and where $d\phi = 0$ was used. Since we cannot use Stokes' theorem directly, we resort to evaluate the last integral over the domain $F_\epsilon$ obtained by subtracting from ${\R}^2$ the closed ball of radius $\epsilon$ with centre at the origin and then going to the limit as $\epsilon \to 0$. We obtain:
\begin{equation} \label{eq17}
\int\limits_F d(\beta \phi) = \lim\limits_{\epsilon \to 0}\int\limits_{F_\epsilon} d(\beta \phi) = \lim\limits_{\epsilon \to 0}\int\limits_{\partial F_\epsilon} \beta \phi=2 \pi \beta(\{0\}).
\end{equation}
Thus, the dislocation current is given by Dirac's delta. To obtain a three-dimensional version of the above,  we consider the same form $\phi=d\theta$ in a cylindrical coordinate system $\rho, \theta, z$ and define the domain $F$ as ${\R}^3$ minus the entire $z$-axis. This layering form can now be pictured as the pages of an open book evenly spread with the spine occupying the $z$-axis. The corresponding dislocation current is now given by:
\begin{equation} \label{eq18}
D[\gamma]= \partial T_\phi [\gamma]= 2 \pi \int\limits_{-\infty}^\infty \gamma_3(0,0,z) dz,
\end{equation}
where $\gamma=\gamma_1 dx+\gamma_2 dy +\gamma_3 dz$ is a 1-form with compact support in ${\R}^3$.

\subsection{A screw dislocation} \label{sec:screw}
Two currents that differ by a closed (i.e., zero-boundary) current have the same boundary. In the context of defects, we may say that two layering currents that differ by defect-free current must exhibit exactly the same defects. This observation can have unexpected physical interpretations. Indeed, let us consider the open-book layering current $T_\phi$ just introduced and let us define the 2-current:
\begin{equation} \label{eq19}
S = T_\phi + a T_{dz},
\end{equation}
where $a$ is a constant. Since $dz$ is a well-defined closed form over ${\R}^3$, so is the associated current $T_{dz}$. Consequently, the boundary of $S$ coincides with the boundary of $T_\phi$. The layering structure corresponding to $S$ consist of applying to the previous `pages' a uniform twist about the $z$-axis. To ascertain that this is indeed the case, notice that, within the domain $F$, $S$ can be regarded as the 1-form $\phi=d\theta +a dz$, a closed form. Locally, therefore, we can write:
\begin{equation} \label{eq20}
\phi= d(\theta +az).
\end{equation}
In other words, locally the submanifolds with equation:
\begin{equation} \label{eq21}
\theta + a z = {\rm constant}
\end{equation}
are integral submanifolds of the distribution generated by $\phi$. These submanifolds describe helicoidal surfaces climbing around the $z$-axis. This {\it screw layering} has a dislocation current identical to that of the open book.

\section{Frank's rules} \label{sec:frank}

Within the classical theory of dislocations in crystals, a prominent role is played by the {\it Burgers vector} concept. Dislocations are assumed to occur along lines only. A {\it Burgers circuit} in an atomic lattice consists of a quadrilateral path situated on an atomic `plane' transverse to the dislocation line and with equal numbers of atomic cell steps on opposite sides. In a perfect crystal, these paths naturally close. The lack of closure (namely, the Burgers vector, denoted by ${\bf b}$), on the other hand, is interpreted as the presence of a dislocation. When, for example, $\bf b$ is parallel to the dislocation line, we have a pure screw dislocation. Clearly, as one advances over the dislocation line, the Burgers vector may change in magnitude and direction, so that the question arises as to whether this change can be arbitrary. Moreover, dislocation lines may meet and branch out, so that a similar question arises in these more involved cases. In an important article \cite{Frank}, F. C. Frank introduced the notion of the {\it law of conservation of Burgers vectors}, formally analogous to Kirchhoff's laws for electrical circuits (charge conservation) or similar laws for fluid flow in pipes (mass conservation). As a consequence of this law, several rules can be deduced. For example, the Burgers vector along a dislocation line must be constant. Moreover, a dislocation line may not end within the crystal, but only at its boundary. At a bifurcation, the vector of the entrant trunk is equal to the sum of the vectors of the outgoing branches. Given the importance of these rules in applications, we want to place them rigorously within the context of the geometrical theory. It will turn out that Frank's rules are various expressions of the general topological criterion that establishes that the boundary operator is nilpotent of degree 2, that is, the boundary of a boundary necessarily vanishes.

\subsection{The constancy rule}
Since Frank's rules deal always with dislocation {\it lines}, we need first to establish the notion of the {\it support of a current} \cite{derham}. A current $T$ is equal to zero in an open set $U$ if $T[\phi]=0$ for all forms $\phi$ compactly supported in $U$. The support of $T$ is defined as the complement of the maximal open set in which $T=0$. Accordingly, we say that a dislocation current $D$ is a dislocation line if its support is a curve. Considering, for specificity, a (three-dimensional) body $\mathcal M$, we investigate the possibility of existence of a curve $L$ within the body, whose ends are not points of $\mathcal M$, with the following properties: (1) $L$ is the support of a dislocation 1-current $D$; (2) $D$ is of the form $T_{uL}$ for some real valued function $u:{\mathcal M} \to {\R}$, namely:
\begin{equation} \label{eq22}
D[\alpha]=\int\limits_L u \alpha,
\end{equation}
with some abuse of notation (in the sense that, within the integral, the 1-form $\alpha$ represents the restriction to $L$ of the original 1-form $\alpha$ with compact support in $\mathcal M$). We will prove that both conditions cannot be satisfied simultaneously unless the scalar function $u$ is actually constant on $L$. The proof starts by remarking that, as a dislocation current, $D$ must be the boundary of some (layering) current $S$:
\begin{equation} \label{eq23}
D=T_{uL}=\partial S.
\end{equation}
Applying the boundary operator, we obtain
\begin{equation} \label{eq24}
\partial D=\partial T_{uL}=\partial\partial S=0.
\end{equation}
Evaluating over a zero-form $f$ yields
\begin{equation} \label{eq25}
\partial T_{uL}[f]=T_{uL}[df]=\int\limits_L uf= \int\limits_L d(uf) - \int\limits_L fdu=- \int\limits_L fdu=0.
\end{equation}
Since $f$ is arbitrary, we conclude that $u={\rm constant}$ on $L$. Thus, if we interpret $u$ as the strength of the dislocation, it follows from this proof that the strength of a line dislocation must be constant. Recall that we are analyzing the dislocations associated with a single layering system, whence the scalar nature of the line dislocation strength.

\subsection{Branching} \label{sec:branching}
If $k$ lines ($L_i, i=1,...,k\ge 3$) meet at a body point $X \in {\mathcal M}$, and if each line is the support of a line dislocation, we have a case of branching. We assume these lines to originate at $X$ and to emerge at the topological boundary of $\mathcal M$ (which is to be seen, as standard continuum mechanics prescribes, as an ordinary differentiable manifold, not as a manifold with boundary). It follows that the boundary of each of the given lines $L_i$ consists of the single point $X$. Denoting by $a_i$ the (constant) strength of the dislocation supported by $L_i$, we associate to the total system the dislocation current $D= \sum\limits_i a_i T_{L_i}$. For an arbitrary compactly supported zero-form $f$, considering that $D$ itself must be a boundary of a (layering) 1-form, we obtain:
\begin{equation} \label{eq26}
0=\partial D[f] = \sum\limits_i a_i \int\limits_{L_i} df = f(X) \; \sum\limits_i a_i .
\end{equation}
Since $f$ is arbitrary, we conclude that:
\begin{equation} \label{eq27}
\sum\limits_i a_i = 0,
\end{equation}
which is Frank's branching rule within the scalar context.

\section{The second integrability criterion} \label{sec:second}

We indicated in section \ref{sec:smooth} that, given a layering form $\omega$, there are two different questions that one may try to answer, each one leading to a different integrability criterion. The first question, which we have been exclusively addressing so far, is whether or not the layers and their respective stacking densities fit well together. The general answer to this question is provided by the closedness of the layering form, namely $d\omega=0$, or, in the singular case, the closedness of the layering current, $\partial T = 0$. The second question, which we have described only in the case of a decomposable\footnote{Note that 1-forms are always decomposable.} layering form $\omega$, is whether or not the associated distribution is completely integrable. According to the theorem of Frobenius, the pertinent condition is the existence of a 1-form $\beta$ such that $d\omega = \beta \wedge \omega$. Clearly, this criterion of integrability is less demanding than the first. Correspondingly, every defect-free decomposable layering form $\omega$ according to the first criterion is also defect-free according to the second, but the converse is not true. Examples are not difficult to construct. In fact, every 1-form in ${\R}^3$ given by the expression $\omega=f(x,y) dx +g(x,y) dy + dz$, for any given smooth functions $f$ and $g$, gives rise to a completely integrable two-dimensional distribution, although $\omega$ is closed only when the cross derivatives $f_{,y}$ and $g_{,x}$ are identical to each other.

\subsection{Coherence at interfaces} \label{sec:coherence}

Working in ${\R}^3$ with natural coordinates $x,y,z$, let $\Sigma$ denote the plane $z=0$. Let, moreover, the upper ($z \ge 0$) and lower ($z<0$) half-spaces be denoted by $H^+$ and $H^-$, respectively. Consider 1-forms $f^+=f^+_1 dx+ f^+_2 dy + f^+_3 dz$ and $f^-=f^-_1dx + f^-_2 dy + f^-_3 dz$ smoothly defined on $H^+$ and $H^-$, respectively, and define a 2-current $T$ by
\begin{equation} \label{eq28}
T[\phi]=\int\limits_{H^+} f^+ \wedge \phi + \int\limits_{H^-} f^- \wedge \phi,
\end{equation}
for arbitrary 2-forms $\phi$ compactly supported in ${\R}^3$. We regard this form as defining a singular 2-dimensional distribution on ${\R}^3$. Its boundary is the 1-current:
\begin{eqnarray} \label{eq29}
\partial T[\psi]= T[d\psi] = \int\limits_{H^+} f^+ \wedge d\psi + \int\limits_{H^-} f^- \wedge d\psi \nonumber \\
    = \int\limits_\Sigma [[f]] \wedge \psi + \int\limits_{H^+} df^+ \wedge \psi + \int\limits_{H^-} df^- \wedge \psi,
\end{eqnarray}
acting on 1-forms $\psi$ with compact support in ${\R}^3$. In this equation, $[[\cdot]]$ denotes the jump operator. Assume now that, for this particular layering current $T$, we want to establish the absence of defects according to our first criterion. Setting
\begin{equation} \label{eq30}
\partial T = 0
\end{equation}
identically, we may first choose arbitrary forms $\psi$ whose support does not intersect $\Sigma$ and obtain the conditions:
\begin{equation} \label{eq31}
df^+ = 0,\;\;\;\;\;\;df^- = 0.
\end{equation}
What these conditions mean is that, as far as the individual layering forms $f^+$ and $f^-$ are concerned, there are no defects at interior points of $H^+$ or $H^-$ and all possible remaining dislocations are concentrated on the surface $\Sigma$. Considering now arbitrary 1-forms $\psi$ whose support does intersect $\Sigma$, we obtain the following extra point-wise conditions on $\Sigma$:
\begin{equation} \label{eq32}
[[f_1]]=0,\;\;\;\;\;\;\;\;[[f_2]]=0.
\end{equation}
The jump $[[f_3]]$ of the $z$-component of the layering form can be prescribed arbitrarily. The geometric interpretation of these results can be gathered by first considering the case in which $f^+_1, f^-_1, f^+_2$ and $f^-_2$ vanish altogether while $f^+_3 = A$ and $f^-_3 = B$, where $A$ and $B$ are different constants. We have then a purely horizontal layering (i.e., parallel to $\Sigma$) that undergoes an abrupt change of density across $\Sigma$. There are no defects in this kind of layering. On the other hand, if $f^+_1, f^-_1, f^+_3$ and $f^-_3$ were to vanish identically while $f^+_2 = A$ and $f^-_2 = B$, the layering would be vertical (i.e., perpendicular to $\Sigma$) and there would be an {\it incoherence} defect across $\Sigma$, unless $A=B$. In a more general case. we would have that the layering consists of leaves that may have a kink upon crossing the surface $\Sigma$, but are otherwise continuous.

We have considered so far the absence of defects according to the first integrability criterion, namely, by checking that the layering current $T$ is closed. We investigate now the consequences of demanding only the existence of a 1-form $\beta$ such that
\begin{equation} \label{eq33}
\partial T= \beta \lrcorner T.
\end{equation}
More explicitly,
\begin{eqnarray} \label{eq34}
\int\limits_\Sigma [[f]] \wedge \psi + \int\limits_{H^+} df^+ \wedge \psi + \int\limits_{H^-} df^- \wedge \psi \nonumber \\  = \int\limits_{H^+} f^+ \wedge(\beta \wedge \psi) + \int\limits_{H^-} f^- \wedge (\beta \wedge \psi).
\end{eqnarray}
Considering first 1-forms $\psi$ whose support does not intersect $\Sigma$, we obtain now the conditions
\begin{equation} \label{eq35}
df^+ = f^+ \wedge \beta,\;\;\;\;\;\;\;\;df^+ = f^+ \wedge \beta.
\end{equation}
In other words, the distributions induced by $f^+$ and $f^-$ in the interior of their respective domains are required only to be completely integrable. By considering 1-forms $\psi$ whose support intersects $\Sigma$, however, we recover the coherence condition (\ref{eq32}). We observe not only that this condition could perhaps be further relaxed, but also that, even in terms of the complete integrability of the upper and lower distributions, a single form $\beta$ should act as integrating factor for both. Clearly, a still weaker form of the Frobenius condition (\ref{eq10}) could be postulated in terms of a current multiplier rather than a smooth form $\beta$. This issue is the subject of further study.

\subsection{Broken leaves} \label{sec:leaves}

We define the 2-current
\begin{equation} \label{eq36}
T[\phi]=\int\limits_\Sigma \phi + \int\limits_{{\R}^3} \alpha \wedge \phi,
\end{equation}
where $\Sigma$ is the same as in the previous example and $\alpha$ is a 1-form defined over ${\R}^3$. Its boundary is:
\begin{equation} \label{eq37}
D[\psi]=\partial T[\psi]= \int\limits_\Sigma d\psi + \int\limits_{{\R}^3} \alpha \wedge d\psi = \int\limits_{{\R}^3} d \alpha \wedge \psi,
\end{equation}
for arbitrary 1-forms $\psi$ with compact support in ${\R}^3$. The stronger integrability criterion $\partial T = 0$ yields the expected condition
\begin{equation} \label{eq38}
d \alpha=0.
\end{equation}
We note that the integral over $\Sigma$ is automatically closed as a current and, consequently, has no effect on the result when applying the (linear) boundary operator. On the other hand, demanding only the satisfaction of Equation (\ref{eq10}), we conclude that a 1-form $\beta$ must exist such that
\begin{equation} \label{eq39}
\partial T [\psi]=(\beta \lrcorner T) [\psi]=\int\limits_\Sigma \beta \wedge \psi + \int\limits_{{\R}^3} \alpha \wedge \beta \wedge \psi.
\end{equation}
This identity implies:
\begin{equation} \label{eq40}
d\alpha= \alpha \wedge \beta \;\;\;\;\;\;\;\;{\rm on}\;{\R}^3,
\end{equation}
and
\begin{equation} \label{eq41}
\beta = 0\;\;\;\;\;\;\;\;{\rm on}\; \Sigma.
\end{equation}
To grasp the meaning of these integrability conditions, we observe that if we were to ignore the integral over $\Sigma$ in the definition of $T$ in Equation (\ref{eq36}), the satisfaction of the integrability condition (\ref{eq40}) would imply that the distribution is completely integrable, thus constituting a regular foliation of ${\R}^3$. The presence of the integral over $\Sigma$ has the effect of breaking the leaves. The vanishing of $\beta$ repairs the damage. An intuitive realization of this picture can be obtained by considering the following sequence of forms:
\begin{equation} \label{eq42}
\alpha_i=(1-c_i) dy + c_i dz,
\end{equation}
where $c_i=c_i(z)$ are scalar functions supported on the interval $[-2^{-i},2^{-i}]$ and such that
\begin{equation} \label{eq43}
\int\limits_{\R} c_i dz = 1.
\end{equation}
The sequence of 2-currents
\begin{equation} \label{eq44}
T_i[\phi]=\int\limits_{{\R}^3} \alpha_i \wedge \phi
\end{equation}
converges to a 2-current $T$ whose leaves are of the broken kind described above.

---------------------------------------------------------------------
\section*{Acknowledgments}
This work has been supported in part by the Natural Sciences and Engineering Research Council of Canada.
\bibliographystyle{unsrt}
\bibliography{mybibfile}
\end{document}